\documentclass{aa}  

\usepackage{epstopdf}
\usepackage{graphicx}
\usepackage[colorlinks=true, linkcolor=blue,  citecolor=blue]{hyperref}% 
\usepackage{txfonts}
\usepackage{color}

\begin{document}

   \title{Very high-energy constraints on the infrared extragalactic background light}

   \titlerunning{Infrared EBL and Cherenkov Observations }

   \author{A. Franceschini\fnmsep\thanks{alberto.franceschini@unipd.it}\inst{1}, L. Foffano\inst{1}\inst{2}, E. Prandini\inst{1}\inst{2},\and
          F. Tavecchio\inst{3}
          }

   \institute{Department of Physics and Astronomy (DFA), University of Padova,
              Vicolo Osservatorio 3, I-35122 Padova, Italy
              %\email{alberto.franceschini@unipd.it}
         \and
             INFN, Istituto Nazionale di Fisica Nucleare (INFN), via Marzolo 8, 35131 Padova, Italy
         \and
             Istituto Nazionale di Astrofisica (INAF), vicolo dell'osservatorio 3, 35131 Padova, Italy
             %\email{c.ptolemy@hipparch.uheaven.space}
             %\thanks{The university of heaven temporarily does not accept e-mails}
             }

   \date{Received April 4, 2019; accepted July 8, 2019}

% \abstract{}{}{}{}{} 
% 5 {} token are mandatory
 
  \abstract
  % context heading (optional)
  % {} leave it empty if necessary  
   {Measurements of the extragalactic background light (EBL) are a fundamental source of information on the collective emission of cosmic sources.}
  % aims heading (mandatory)
   {At infrared wavelengths, however, these measurements are precluded by the overwhelming dominance from interplanetary dust emission and the Galactic infrared foreground. Only at $\lambda > 300 \ \mu$m, where the foregrounds are minimal, has the infrared EBL (IR EBL) been inferred from analysis of the COBE maps. The present paper aims to assess the possibility of evaluating the IR EBL from a few micrometers up to the peak of the emission at >100 $\mu$m using an indirect method that avoids the foreground problem.  }
  % methods heading (mandatory)
   {To this purpose we exploit the effect of pair-production from gamma-gamma interaction by considering the highest-energy photons emitted by extragalactic sources and their interaction with the IR EBL photons. We simulate observations of a variety of low-redshift emitters with the forthcoming Imaging Atmospheric Cherenkov Telescope (IACT) arrays (CTA in particular) and water Cherenkov observatories (LHAASO, HAWC, SWGO) to assess their suitability to constrain the EBL at such long wavelengths.}
  % results heading (mandatory)
   {We find that even under the most extremely favorable conditions of huge emission flares, extremely high-energy emitting {\it blazars} are not very useful for our purpose because they are much too distant  (>100 Mpc the nearest ones, MKN 501 and MKN 421). Observations of more local  AGNs displaying very high-energy emission, like low-redshift radio galaxies (M87, IC 310, Centaurus A), are better suited and will potentially allow us to constrain the EBL up to $\lambda \simeq 100\ \mu$m.
}
  % conclusions heading (optional), leave it empty if necessary 
   {}

   \keywords{giant planet formation --
                stability of gas spheres
               }

   \maketitle
%
%-------------------------------------------------------------------

\section{Introduction}

Measurements of the extragalactic background light (EBL) are a fundamental source of information on the collective emission of all cosmic sources from the Big Bang to today. As such, the EBL is not only an important radiative constituent of the local universe, but also offers critical constraints on all astrophysical and cosmological processes taking place during the formation of cosmic structure \citep{longair}.

Unfortunately, the only spectral region where its direct measurement has been possible is the sub-millimeter band ($200<\lambda<900\ \mu$m), where the local foreground emissions are minimal. There, the EBL was identified and measured from data obtained with the FIRAS instrument on-board the COsmic Background Explorer (COBE) \citep{puget, fixten, hauser, lagache99, hauseranddwek}.

At all shorter wavelengths, however, these measurements are complicated by the overwhelming dominance of local emission from both the Galaxy and the solar system.  These foregrounds can be orders of magnitude more intense than the expected levels of the EBL, meaning that their evaluation is subject to large uncertainties, or is even virtually impossible. 

For this reason, during the last two or three decades an important effort has been dedicated to attempt empirical determinations of the EBL with an independent, though indirect, method that takes advantage of the physical interaction between very high-energy (VHE) photons emitted by distant active galactic nuclei (AGNs, the \textit{blazar} population in particular) and those of the much lower-energy EBL, with the consequent $e^+ e^-$ pair-production. From a determination of the corresponding spectral cutoff in the background {blazars}, constraints on the EBL intensity have been inferred  \citep[][henceforth AF2008, among many others]{Stecker, Stanev, Aharonian2006, AF2008}. Performing the same exercise on a sample of objects at different redshifts can also allow us to map the time evolution of the cosmic photon number density, and thereby constrain the history of the EBL photon production.
The latter has been applied successfully in mapping the EBL evolution and galaxy star-formation history based on the ten-year Fermi satellite observations of the blazar high-energy (HE) flux \citep{FermiLAT}. Thanks to the Fermi HE spectral coverage, this analysis constrained the EBL intensity and time evolution in the UV-optical ($0.1<\lambda<3\ \mu$m).

Over the wide wavelength interval from 3 to 300 $\mu$m, direct observations of the infrared (IR) EBL are precluded by the overwhelming dominance of the interplanetary dust emission and the IR foreground from Galactic dust \citep{hauseranddwek}. This is unfortunate because deep observations with IR space- and ground-based telescopes have revealed that this spectral region is very rich in astrophysical and cosmological information.  Many very luminous sources at high redshifts have been detected and identified, which are interpreted as likely tracing major episodes of the formation of galaxies, AGNs, and quasars \citep{blain, AF2001, madau}.

Infrared telescopes can easily detect point sources, but cannot identify more diffuse emission, like extended halos of dust emission or truly diffuse processes, because of the huge background noise. Also, because of the source confusion problem due to the limited angular resolution at such long wavelengths, faint sources are out of reach of the existing IR observatories.

For all these reasons, measuring the total IR EBL between 3 and 300 $\mu$m would offer a valuable contribution to extragalactic astrophysics and cosmology.
The present paper aims at critically assessing the possibility of constraining the EBL from the gamma-gamma interaction by considering the highest-energy photons emitted by extragalactic sources and their interaction with IR EBL photons.
We simulate observations of a variety of low-redshift emitters with forthcoming Cherenkov Telescope Arrays (CTA and others) to assess their suitability for constraining the EBL at such long wavelengths.

We determine whether extremely high-frequency peaked blazars will be useful to our purpose, considering that they tend to be distant objects 
(with the nearest ones, MKN 501 and MKN 421, having distances >100 Mpc), or if observations of more local VHE-emitting AGNs, such as low-redshift radio galaxies (M87, Centaurus A), might be better suited to constraining the EBL up to $\lambda \simeq 100\ \mu$m.
 
The paper is organized as follows.
Section 2 reviews the astrophysical significance of the IR EBL and motivations for its estimation.
%\LEt{Please check that I have retained your intended meaning. YES}. 
Section 3 considers possible targets of VHE observations, such as extreme \textit{blazars} at low redshifts.
Section 4 examines how to take advantage of blazar flaring states to optimize the analysis.
Section 5 expands the simulation to the most local and luminous VHE-emitting AGNs and reviews the performances of the best-suited facilities (CTA, HAWC, SWGO, LHAASO).
Section 6 summarizes our results.

\noindent
We adopt a standard cosmological model with $H_0=70$ km s$^{-1}$ Mpc$^{-1}$, $\Omega_m=0.3$, $\Omega_\Lambda=0.7$.

%--------------------------------------------------------------------
\section{The IR extragalactic background light and its astrophysical significance}

Far-infrared (FIR) observations of the extragalactic sky, starting from the IRAS  survey \citep{Soifer}, and then exploiting the sequence of space IR telescopes, ISO, Spitzer, Herschel, Planck, together with ground-based facilities (SCUBA/JCMT, IRAM, APEX, ALMA, among others) have offered a new, unexpected view of the universe.
These showed that major phases of both star formation in massive galaxies and gravitational accretion in AGNs occur inside heavily dust-enshrouded clouds and media \citep{AF2001, AF2010, blain, Gruppioni, eales}.

Concerning star-formation, it is natural to expect that the gravitational collapse of the parent cloud, leading to the formation of a star cluster, would be favored by the attenuation of the environmental ionizing radiation field by an embedding dust cocoon. This dust absorbs the UV flux from young stars and re-radiates thermally in the FIR.
Large reservoirs of gas and dust surrounding the circum-nuclear regions 
also explain the large FIR emission of all the radio-quiet AGNs \citep{Antonucci, Fritz}. In type-2 AGNs, which are interpreted as the sub-class for which the observer's line-of-sight intercepts the dusty torus, the mid-infrared (MIR) and FIR portion of the spectrum dominates all other emissions.

\subsection{Infrared extragalactic background light from known sources}

\begin{figure}
   \centering
   \includegraphics[width=10cm]{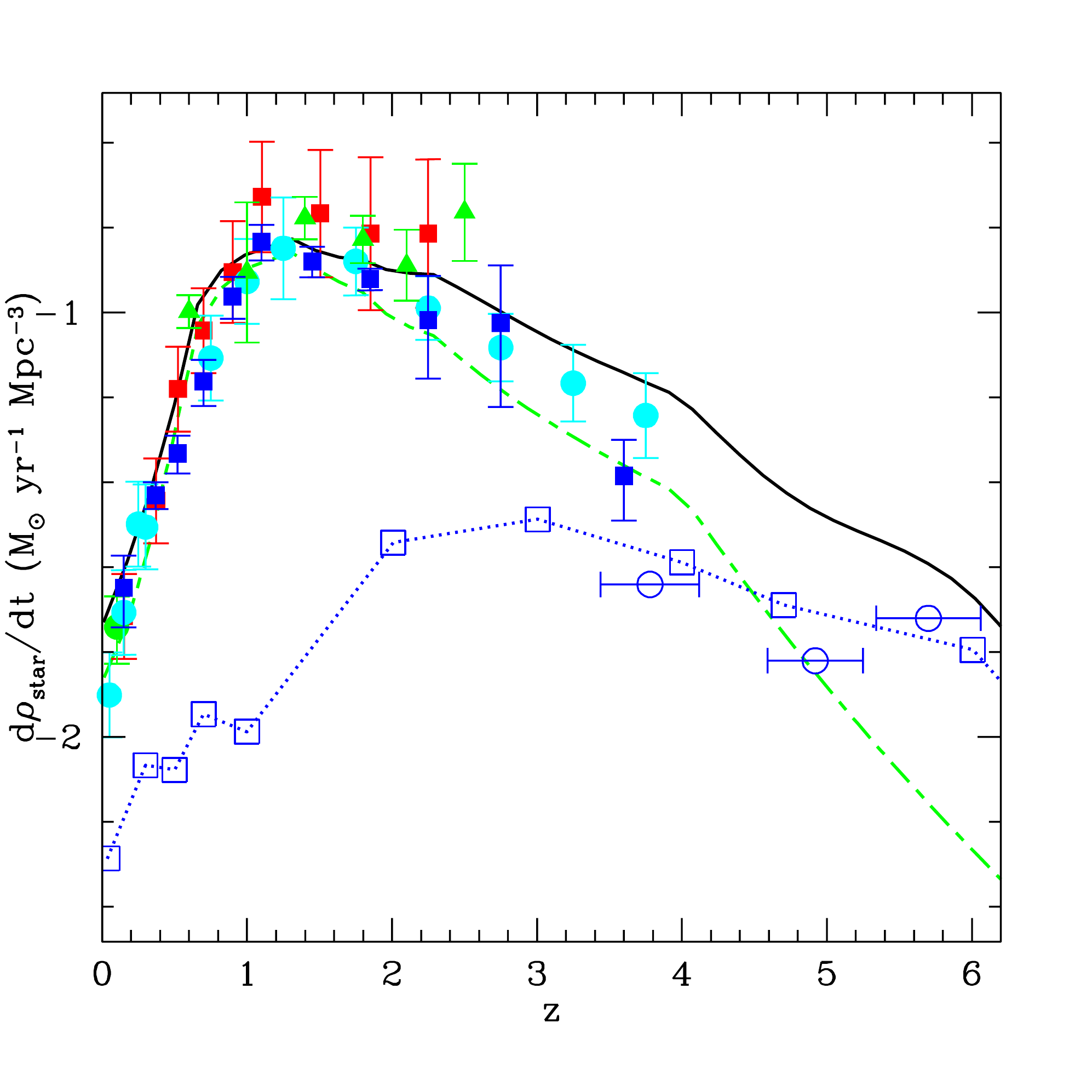}
      \caption{
Global SFR per comoving volume traced by IR and UV observations as a function of redshift. The plot adopts a standard Salpeter stellar IMF.
Here the blue filled squares are an estimate of the SFR density by \cite{Gruppioni} after subtraction of the type-I AGN contribution, which should be matched to the green long-short dash curve that is the total contribution to SFR by IR-selected LIRGs and ULIRGs (type-I AGNs excluded).    
The blue dotted line with open squares traces our estimated comoving SFR density derived from UV observations, with no extinction corrections. 
The black continuous curve shows the total SF activity by FIR and far-UV selections.
Open circle datapoints are from \cite{Giavalisco}, also uncorrected for dust-extinction.
}
\label{sfr3}  
   \end{figure}

An illustration of the relevance of dust absorption and thermal re-emission in galaxies during the Hubble time is reported in Figure \ref{sfr3}.  This shows that from approximately the first two billion years after the Big Bang up until the present time the majority of the radiant power emitted by young stars, tracing the star-formation rate (SFR) in the universe, is absorbed by dust and re-emitted thermally in the MIR and FIR.
This figure is based on the analysis by \citet[][henceforth FR2017]{AF2017}  of a vast amount of multi-wavelength data on the galaxy and AGN high-redshift emissivity from a few  to 1000 $\mu$m: details of this study can be found in this latter paper and in \cite{AF2010}.
The comoving SFR density inferred from IR data is shown as the green dashed line. 
In order to analyze the UV-optically selected galaxies and their contribution to the SFR history, luminosity function data by %\LEt{Semicolons are used to separate references in parentheses only. Please use regular punctuation when including references in the text; that is, commas and and. Some help is provided by the Journal at https://www.aanda.org/author-information/latex-issues/references.} 
 \cite{Arnouts}, \cite{Schiminovich},   \cite{Wyder},  \cite{Shimazaku}, and  \cite{Cucciati} are used. 
The blue dotted line with open squares in Fig. \ref{sfr3} reproduces the SFR density that we estimate by converting the far-UV luminosities with the factor appropriate for a standard \cite{Salpeter} initial mass function  %\LEt{Please merge these two parentheses using a semicolon to separate their contents.} 
\citep[IMF,][]{Kennicutt}.

%
%                                                One column figure
%----------------------------------------------------------------- 

%-----------------------------------------------------------------
%
%                                                One column figure
%----------------------------------------------------------------- 
%   \begin{figure}
%   \centering
%   \includegraphics[width=8cm]{AV.eps}
%      \caption{
%The global averaged extinction of star-formation in galaxies as a function of redshift. 
%The comoving SFR densities $SFR_{UV}$ and $SFR_{IR}$ are those reported in Fig. \ref{sfr3}. 
%              }
%\label{AV} % macro AV (SP.MON)
%   \end{figure}
%-----------------------------------------------------------------

The IR and UV curves in Fig. \ref{sfr3} show a very different evolution. 
Ultraviolet galaxies dominate the global SFR at redshifts $z>4$, but become almost negligible at lower redshifts.  Dust-obscured IR galaxies are the largest contributors to the star-formation activity at $z<3$.

The trends in Fig. \ref{sfr3} have a natural interpretation if we consider that a concentration of dust large enough to produce significant extinction is released to the galactic interstellar medium on a time-scale of only a few gigayears following the first-light epoch.   This implies that UV-bright objects in typically low-mass proto-galaxies dominate the early, mostly dust-free phases, while at lower redshift the bulk of the star formation activity happens inside dust-obscured media.  
At $z<1$ the IR emissivity shows a fast decay, tracing a decrease of the SFR and of the dust-rich diffuse gas in galaxies, with a corresponding reduction of the dust opacity and IR re-radiation. 
The shallower decrease of the UV-bright part at $z<2$ is because the general decrease of the SFR density is partly compensated by the reduction of the average galactic opacity.

In the following, we adopt as our reference EBL model that described in FR2017 and AF2008.  This estimate is to be considered as a lower boundary to the real EBL as it is based on only discrete identified sources.

\subsection{ Uncertainties on the extragalactic background light model and further possible contributions by diffuse components}

There are various reasons why the real total EBL intensity might be significantly different from the FR2017 expectations. 
The first one is the possible occurrence of low surface-brightness IR emission not detected by the moderately sensitive FIR telescopes. 
This might be produced, for example, by a distribution of dust particles in halos around galaxies due to the nucleo-synthetic processes in %\LEt{please use "Galactic" (and "Galaxy") when referring to the Milky Way; please edit for this throughout.}
 Galactic stars and to mass loss from galaxies driven by supernova explosions or collective stellar winds.
These outflows have indeed been observed in galaxy clusters, where the stellar products are found as heavy elements polluting the intra-cluster medium (ICM; although in this case the formation of dust grains may be prevented by the hot intra-cluster (IC) plasma (e.g. \citealt{Sarazin}; see also observational constraints in \citealt{GL}). 
The amount of the X-ray emitting IC plasma and its metallicity imply that about two times the metal content in stars is lost from galaxies. If a similar phenomenon characterizes the numerous field galaxies, this amount of metals ejected from galaxies would be free to coalesce into dust grains, be illuminated by a weak radiation field from the parent galaxy, and then emit at IR wavelengths.

Another source of low surface-brightness emission, including diffuse intra-halo light from stars, has been suggested by \cite{Zemcov14} to explain excess near-infrared (NIR) intensity fluctuations measured with the CIBER rocket telescope, \citep{Zemcov13}.
Any dust associated to these diffuse stars would produce an IR EBL excess signal in the FIR not individually detectable by IR telescopes. Indeed, significant correlations between background fluctuations in the NIR and FIR from \textit{Spitzer} and \textit{Herschel} observations are reported by \citet{Thacker}.

The modest resolution and sensitivity of the latter also implies that only the high-luminosity part of the distant source population is detectable.
The question is then how much of the cosmic IR emissivity is hidden in such faint objects that are not accounted for by IR observations in FR2017. 
Insight into the uncertainty related to the extrapolation of IR luminosity functions may be obtained considering very deep \textit{Spitzer} 24 $\mu$m photometric imaging of optical high-z sources by %\LEt{Semicolons are used to separate references in parentheses only. Please use regular punctuation when including references in the text; that is, commas and and. Some help is provided by the Journal at https://www.aanda.org/author-information/latex-issues/references.} 
\cite{Daddi}, \cite{Daddi2007}, and \cite{Santini}. A tentative conclusion of these analyses seems to be that the bulk of source emissivity occurs in relatively massive galaxies, around or slightly below the characteristic stellar mass ($10^{11}\ M_\odot$), and only moderate fractions (10-20\%) of it come from faint low-mass galaxies unreachable by current IR observations. This is of course very tentative and uncertain, awaiting systematic studies at very faint fluxes, for example by the ALMA interferometer.

Another possible source of excess EBL signal might be a truly diffuse background, for example due to decaying or annihilating particles. If falling in our relevant IR wavelength range, the signal would correspond to rest-mass energies of such hypothetical particles ranging from 0.005 to 0.5 eV. 
One may speculate that dark matter particles of such masses would produce a rather smooth IR background, undetectable by present instrumentation.

Let us finally stress that the whole interval from 24 to 70 $\mu$m  was not covered by any of the space missions during the last 30 years, and this is reflected in the lack of data on the IR EBL within this wavelength interval.
Attempts, as discussed in the following, to constrain the total EBL intensity at IR wavelengths - 5 to 300 $\mu$m - via an indirect approach sensitive to the total light intensity are therefore well motivated.

\section{Relevant forthcoming and future VHE facilities}

Several future facilities are dedicated to expanding our observational capabilities for the detection of the highest photon energies above 10 TeV, including %\LEt{Please spell out all acronyms the first time they appear in the paper, followed by the abbreviation in parentheses, both in the abstract and again in the main text. After that, please only use the abbreviation. See A and A language guide Section 5.2.4 www.aanda.org/language-editing}
Imaging Air Cherenkov Telescopes (IACT) and the High-Altitude Water Cherenkov observatories. 
We review in this section the most promising of these facilities, to be considered in our later analysis.

The most important project in this sense is the Cherenkov Telescope Array (CTA), which will consist of two large arrays: one in the north (CTA-N, $28^\circ$ degrees north), one in the south (CTA-S, $24^\circ$ south).
Sensitivity performances for CTA are reported in \url{https://www.cta-observatory.org/science/cta-performance/}. As already mentioned, our analysis adopts the southern CTA sensitivity, which will include 70 widely distributed small telescopes optimized for the detection of the highest-energy photons. In doing so, we assume that even positive-declination sources might be observable at high Zenith angles. The northern array will be four times less sensitive at such high energies.

Indeed, high-zenith-angle IACT observations of VHE photons (above 10 TeV) offer a remarkable advantage in terms of atmospheric collection area, because Cherenkov photons propagate through a thicker atmosphere over a longer path. Of course, the larger atmospheric collecting area implies better sensitivities for $>10$ TeV photon detection by IACT, but requires high-zenith observations of the source %integrations close to %\textcolor[rgb]{0.984314,0.00784314,0.027451}{ source's rise and set above the horizon\LEt{the meaning here is unclear; please reformulate so as not to use the possessive apostrophe.}}
, accurate monitoring of the atmosphere transparency, proper calibration of the Cherenkov signals, and cameras sensitive in the red (R to I bands). A detailed program of observations and Monte Carlo simulations is currently in progress at the MAGIC observatory to characterize and validate the high-zenith-angle observation technique \citep{Mirzoyan}.

In the following, we consider CTA sensitivity predictions  corresponding to two different integration times and flux confidence limits. The first prediction assumes a total integration time of 50 hours, and a limiting flux in every photon energy bin of $5\sigma$. This is a very conservative assumption when applied to every individual bin: for this reason we also considered a 2$\sigma$ limit per energy bin (still corresponding to a 95\% confidence), and an increased  total integration to 100 hours on-source.
We note that the signal-to-noise ratio for such high-energy Cherenkov observations scales roughly linearly with the integration time, because we can assume these will essentially be performed in a no-background noise condition.
This behavior is confirmed by simulated data in the CTA science page.
Altogether, the 100-hour 2$\sigma$ limit is a factor five deeper than the 50-hour 5$\sigma$ limit. 
These sensitivity limits are shown in Figs. 4 to 8 below as black open squares centered in each energy bin and connected by continuous lines.

Thin dotted lines in the figures  
denote the five-year detection limit of the future large Southern Gamma-Ray Survey Observatory %\LEt{Please spell out all acronyms the \textit{first} time they appear in the paper, followed by the abbreviation in parentheses, both in the abstract and again in the main text. After that, please only use the abbreviation. See A and A language guide Section 5.2.4 www.aanda.org/language-editing}
(SWGO) that is currently under consideration 
\citep{Albert19}. 
The five-year sensitivity limits of the HAWC array \citep{DeYoung} are reported as blue dashed lines, but these do not appear to offer competitive performances with respect to long integrations with the other Cherenkov telescopes.

An excellent opportunity to cover the extreme-energy regime in the  future (completion of the installation expected by the end of 2021) will be offered by the LHAASO experiment \citep[$29^\circ$ north, ][]{DiSciascio}. This will consist of a 1 km$^2$ scintillator array, a closely packed surface water Cherenkov detector facility with a total area of about 78,000 m$^2$, and an array of wide-field-of-view air Cherenkov telescopes.  
The red short dashed lines in the figures correspond to assuming 5 years of continuous monitoring of the north sky by this observatory.

There is an important difference between the long observational campaigns performed with either IACT and water Cherenkov instruments. To achieve sufficient sensitivity, the latter require observations of 1 year or more, while CTA will achieve them on timescales of tens of hours. In the presence of variability, which is almost the norm at such high photon energies, dedicated campaigns with CTA are preferred.
The slow photon accumulation rates of the water observatories require integration of the source fluxes during different emission phases, with likely spectral variations accompanying those in flux. The interpretation of such integrated spectra might be complicated and would require careful data editing.

\section{Infrared extragalactic background light  constraints from Cherenkov observations of very low-redshift extreme blazars}

In this and the following sections we discuss how to obtain constraints on the IR EBL intensity via Cherenkov observations of VHE-emitting sources. This is achieved by matching the observed VHE source spectra with theoretical spectra including the exponential cutoff originating from photon absorption and pair production from gamma-gamma interactions.
We assume as a guideline in our simulated analysis that the IR EBL intensity is exactly that modeled by FR2017, which includes only the known source contributions. We then discuss which category of VHE-emitting sources are best suited to constraining the EBL at $\lambda > 10\ \mu$m.

Our calculation of the photon--photon opacities follow our approach discussed in AF2008 and FR2017.

When looking for the EBL signals at the longest IR wavelengths, we need to analyze VHE spectra at the highest photon energies. 
As a guideline, the maximum of the cross section for pair production between interacting photons is given by the relation \footnote{
As discussed in \citet[][see also AF2008]{Heitler}, the cross-section for photon--photon interaction has a maximum when the product of photon energies is $\epsilon_1 \epsilon_2\simeq 4m_e^2c^4/[1+cos(\theta)]$, $\theta$ being the photon collision angle. For an average $\theta\simeq \pi/2$, and transforming the energy $\epsilon_1$ to photon wavelength $\lambda_{max}$, one obtains Eq. \ref{energy}.
}:
\begin{equation}
\lambda_{max} \simeq 1.24 (E_\gamma [TeV])\; \mu m ,
\label{energy}
\end{equation}
where $E_\gamma$ is the VHE or high-energy background photon in TeV and $\lambda_{max}$ is the wavelength of the foreground EBL photons. 
Constraining EBL photon wavelengths above 10 $\mu$m requires measurements of the VHE photons above 10 TeV.
Such high-energy photons are subject to strong opacity effects that are a quickly increasing function of the background source distance.

To compensate for that, we first consider the inclusion of cosmic sources able to produce the highest-energy gamma-ray photons. A clear choice is {blazars}  of the {extreme high-frequency peaked} BL-Lac (EHBL) class. In the blazar classification scheme, a sequence ranging from luminous and distant flat-spectrum radio quasars to low- (LBL), intermediate- (IBL) and high-frequency (HBL)  peaked sources exists. There appears to be an anti-correlation between the average source luminosity and the peak frequencies of the synchrotron and inverse Compton components in these sources. The EHBL \textit{blazars} constitute the extreme of this sequence as the AGNs displaying the highest peak frequencies, and are therefore potentially the best sources of the highest-energy photons.

Characterized by a synchrotron emission component peaking above the frequency of $10^{17}$ Hz, this extreme population has been studied by several teams \citep[][among others]{Costamante,Tavecchio09,Tavecchio10}. A physical interpretation of this extreme phenomenon considered either inverse Compton in the deep Klein-Nishina regime (with a relatively large Doppler factor $\delta=[\Gamma_{Lorentz}(1-\beta cos\theta)]^{-1}$, $\Gamma_{Lorentz}$ being the source Lorentz factor in the observer's frame and $\theta$ the jet to observer angle) in the context of leptonic models
\citep[e.g. the synchrotron self-Compton model,][]{Tavecchio10}, or hadronic emission processes \citep{cerruti}.

In the following analysis we make use of our EHBL selection and analysis work \citep{Foffano} based on the \textit{Swift}-BAT hard X-ray and the \textit{Fermi}-LAT gamma-ray surveys, including 32 candidate objects. 
We note that the sample mostly includes sources observed during quiescent phases and excludes flaring states.
 
The analysis by \citet{Foffano} has revealed a dichotomy in the sub-sample of the TeV detected objects during quiescent phases, some having fast-rising spectra (once corrected for EBL absorption) up to 10 TeV (\textit{hard} EHBL), some others with converging spectra already at 1 TeV and below (\textit{soft} EHBL).  Six of the sample of 32 sources do not have the redshift and cannot be used for our analysis, while all of the 5 identified {hard} EHBLs display relatively high redshifts ($0.12<z<0.19$) and cannot be used. 
In particular, the prototypical object of this class, 1ES 0229+200 (see Fig. 5 in \citealt{Foffano}) has been detected by HESS up to an energy of about 10 TeV \citep{Aharonian2007}, but the relatively high redshift of the source (z=0.139) implies a fast drop-off above that energy, making it unsuitable for our analysis.

%-----------------------------------------------------------------
%                                                One column figure
%----------------------------------------------------------------- 
   \begin{figure}
   \centering
   \includegraphics[width=10cm]{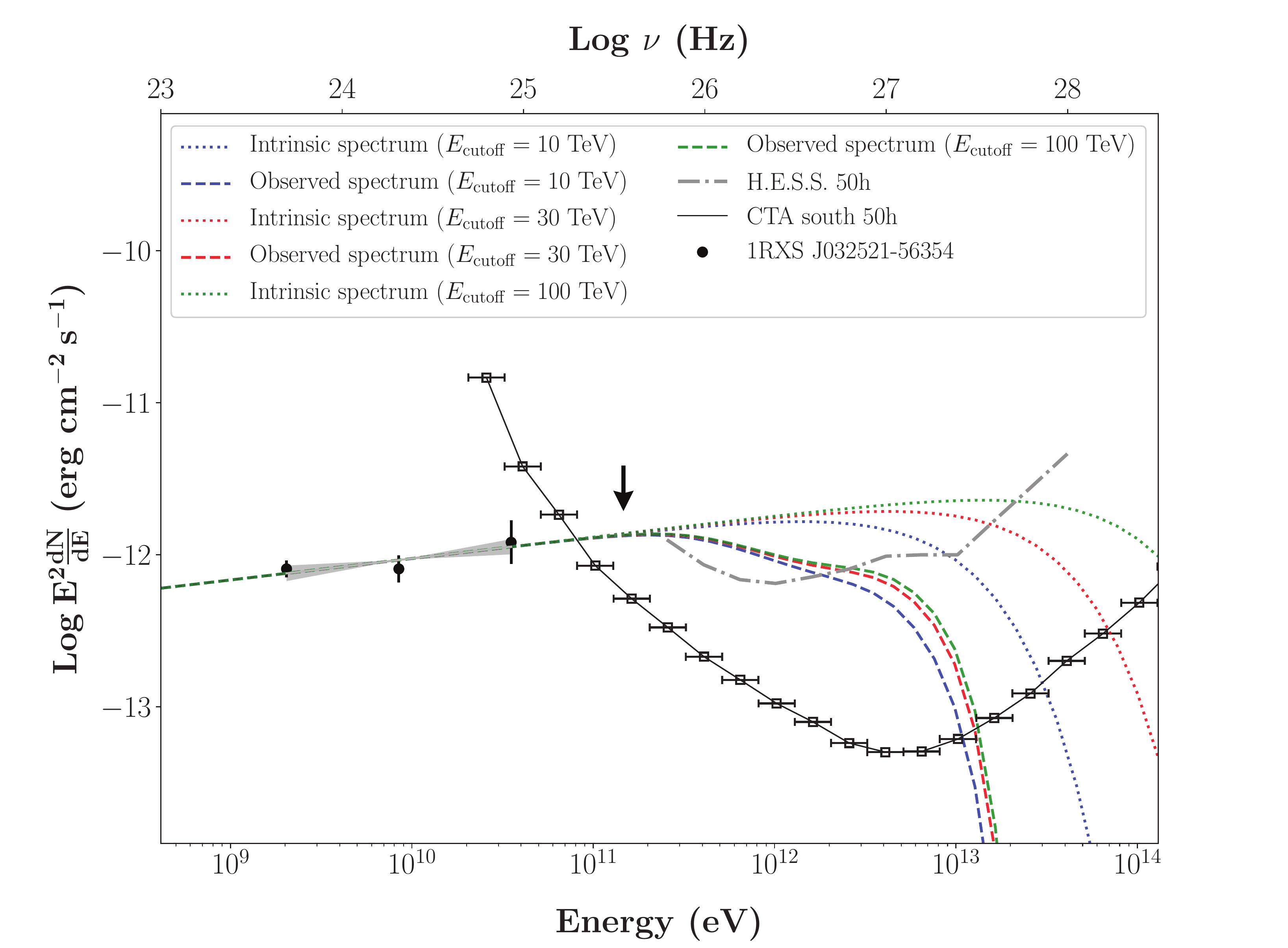}
      \caption{
Extrapolation to the very high energies of the spectrum of the EHBL object 1RXS J032521.8-56354 at z=0.06, assuming an intrinsic power-law spectrum (dotted line) with intrinsic exponential cutoff energies at 10, 30, and 100 TeV (the three dotted lines, see text). The short-dashed lines are the corresponding EBL-absorbed spectra from FR2017. The CTA-south 50h sensitivity is also shown in energy bins.
              }
\label{0325} 
   \end{figure}
%-----------------------------------------------------------------
%-----------------------------------------------------------------
%                                                One column figure
%----------------------------------------------------------------- 
   \begin{figure}
   \centering

    \includegraphics[width=10cm]{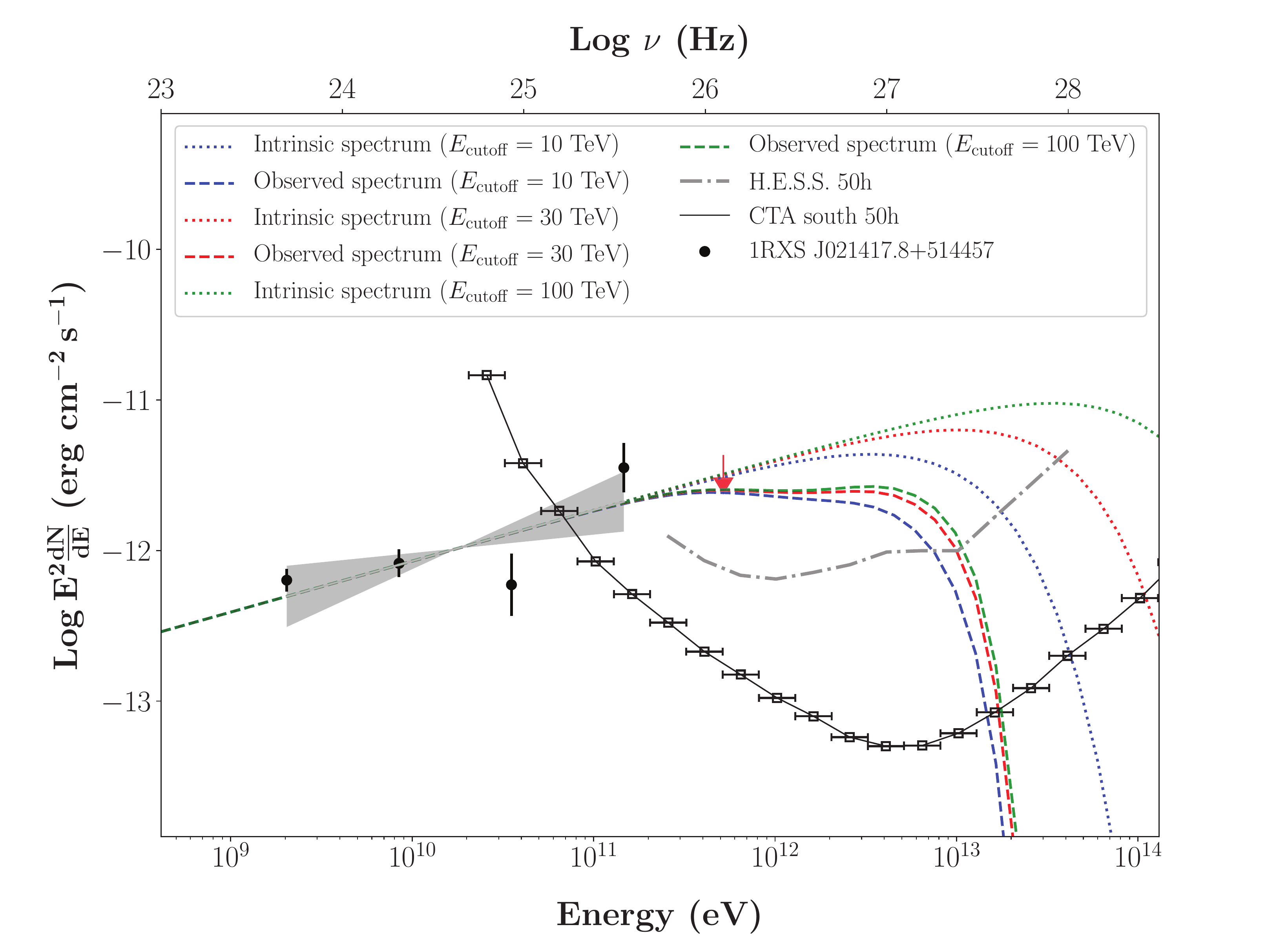}
      \caption{
Extrapolation to the very high energies of the spectrum of the EHBL object 1RXS J021417.8+514457 at z=0.049, assuming an intrinsic power-law spectrum (dotted line) with cutoff energies and absorption corrections as in Fig. \ref{0325}. 
              }
\label{0214} 
   \end{figure}
%-----------------------------------------------------------------
%                                                One column figure

Among the remaining sources in the Foffano et al. paper, we consider those at the lowest redshifts, 1RXS~J032521.8-56354 at z=0.06 and 1RXS~J021417.8+514457 at z=0.049. Because neither is detected with Cherenkov observatories (the former source was observed but was not detected by HESS), we have extrapolated the Fermi-LAT points to the VHE by assuming power-law spectra with exponential cut-off,
$dN/dE\propto E^{-\Gamma}exp(-E/E_{cutoff})$. While the power-law function describes the rising part of the high-energy peak (the hard-TeV EHBLs show power-law spectra up to the deep TeV range; see also \citealt{Costamante18}), the cut-off at energies above 10 TeV reproduces the possible peak of such high-energy emission. To be as general as possible, in Figures \ref{0325} and \ref{0214}  we assume three different values for $E_{cutoff} : $ 10, 30, and 100 TeV. We have verified that this assumption produces spectra in full agreement with the published upper limits by current IACT observations. All details about the observational data on these sources are reported in \cite{Foffano}.

We then correct these intrinsic spectra by EBL absorption, given the source redshift, according to FR2017. Intrinsic and EBL-absorbed spectra are reported in Figs. \ref{0325} and \ref{0214} as the dotted and short-dashed lines, and matched to the CTA expected sensitivity for 50 hours of observations. 
The figures show that even for such long integrations with the new generation of observatories, these EHBL sources would not be detected significantly above 10 TeV, which, from eq. \ref{energy}, is the minimum energy we would need to estimate the IR EBL.
We note that this is a rather robust conclusion and is essentially independent from our adopted detailed intrinsic shape of the sources, because the expected spectrum is largely determined by the EBL absorption, unless the intrinsic source spectrum has a cut-off at much lower photon energies.

Clearly, EHBL {blazars} observed in the quiescent state, as considered so far, are too faint and extinguished to allow constraint of the IR EBL longward of 10 $\mu$m.

\section{Taking advantage of HBL flaring states}

Some progress would certainly be achieved by observations of much more luminous {blazars}, such as nearby HBL objects, observed during the most extreme flaring events. 

\begin{figure}
   \centering
   \includegraphics[width=9cm]{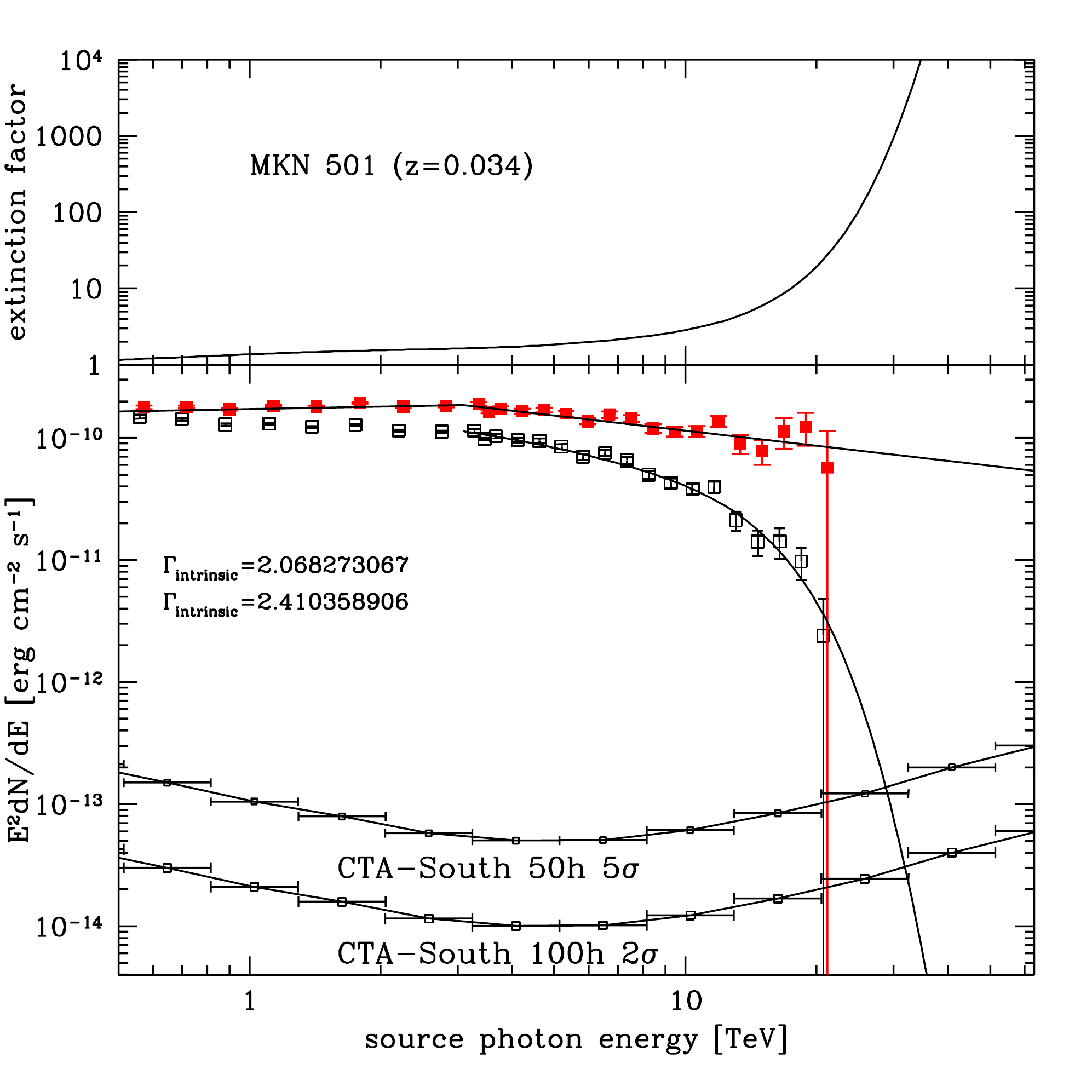}
\caption{
\textit{Top panel: } Photon--photon absorption correction ($exp[\tau_{\gamma\gamma}]$) for the source MKN 501 at $z=0.034$ based on the EBL model by FR2017.
\textit{Bottom:} Observed (open black) and absorption-corrected (filled red) spectral data. Data are taken from \cite{Aharonian2001}; see also \cite{Aharonian1999}. The intrinsic spectrum is fitted by a broken power-law with relatively flat photon spectral indices, as indicated in the insert.
The energy-binned sensitivity limits for 50-hour $5\sigma$ and 100-hour $2\sigma$ of CTA South are also shown.
}               
\label{MKN501} % macro AV (SP.MON)
   \end{figure}

One such occurrence for the close-by blazar MKN 501 is reported in Fig. \ref{MKN501}. These data correspond to the famous outburst registered for the source in 1997 and observed with HEGRA \citep{Aharonian1999}, during which the source flux exceeded that of the Crab and had a very hard spectrum with no sign of convergence up to the highest-energy detected photons of $\sim 20$ TeV. That was an historical event that has not been repeated on a timescale of 30 years. A similarly hard spectral behavior was observed during a flare in 2014 detected by HESS \citep{Abdalla2019}, although with a flux lower than that of the 1997 event.

%-----------------------------------------------------------------
%                                                One column figure
%----------------------------------------------------------------- 
%-----------------------------------------------------------------

The planned CTA-south sensitivity for a 50-hour observation will be three orders of magnitude deeper than the source intrinsic flux and with sensitivity higher than the HEGRA by a factor of approximately 30. In spite of such ideal conditions, CTA would be expected to detect the source only up to 30 TeV at best. 
The reason for this modest improvement stems from the extremely steep exponentially converging spectrum at the highest energies due to EBL absorption. One may gain orders of magnitude in sensitivity compared to current instrumentation, but the increase in spectral coverage turns out to be very marginal.
Similar conclusions would apply for the other well-known low-redshift blazar, MKN 421, in this case the situation being slightly less favorable because of its softer VHE spectrum (see e.g., FR2017).

In conclusion, {blazars} of all classes, even the closest ones in the most favorable conditions of an exceptionally excited state, fall short of providing us with sufficiently high-energy photons to constrain the most interesting portion of the IR EBL, which is above a few tens of micrometers where the bulk of its energy resides.
Very high-energy-emitting {blazars}, in spite of being the most prominent sources of VHE photons, are too rare to include sufficiently nearby objects to make the absorption effects valuable %effective%\LEt{please reword this sentence so as to avoid "effect effective".}
 for long-IR-wavelength photons.

\section{Observing the most local VHE-emitting AGNs}

It is clear from our previous analysis that, independently of the source luminosities, an essential parameter to constrain the IR EBL is the source distance. We therefore expanded our analysis to include the closest VHE-emitting AGNs. 
A much more numerous population than the blazars are the radio galaxies, which according to the unification scheme are their parent population \cite{UrryPadovani}.
Indeed, several radio galaxies were already detected by current IACTs up to the highest energies, often displaying relatively hard spectra. Thanks to their relatively high number density, some populate the very local universe.

%-----------------------------------------------------------------
%                                                One column figure
%----------------------------------------------------------------- 
   \begin{figure}
   \centering
   \includegraphics[width=9cm]{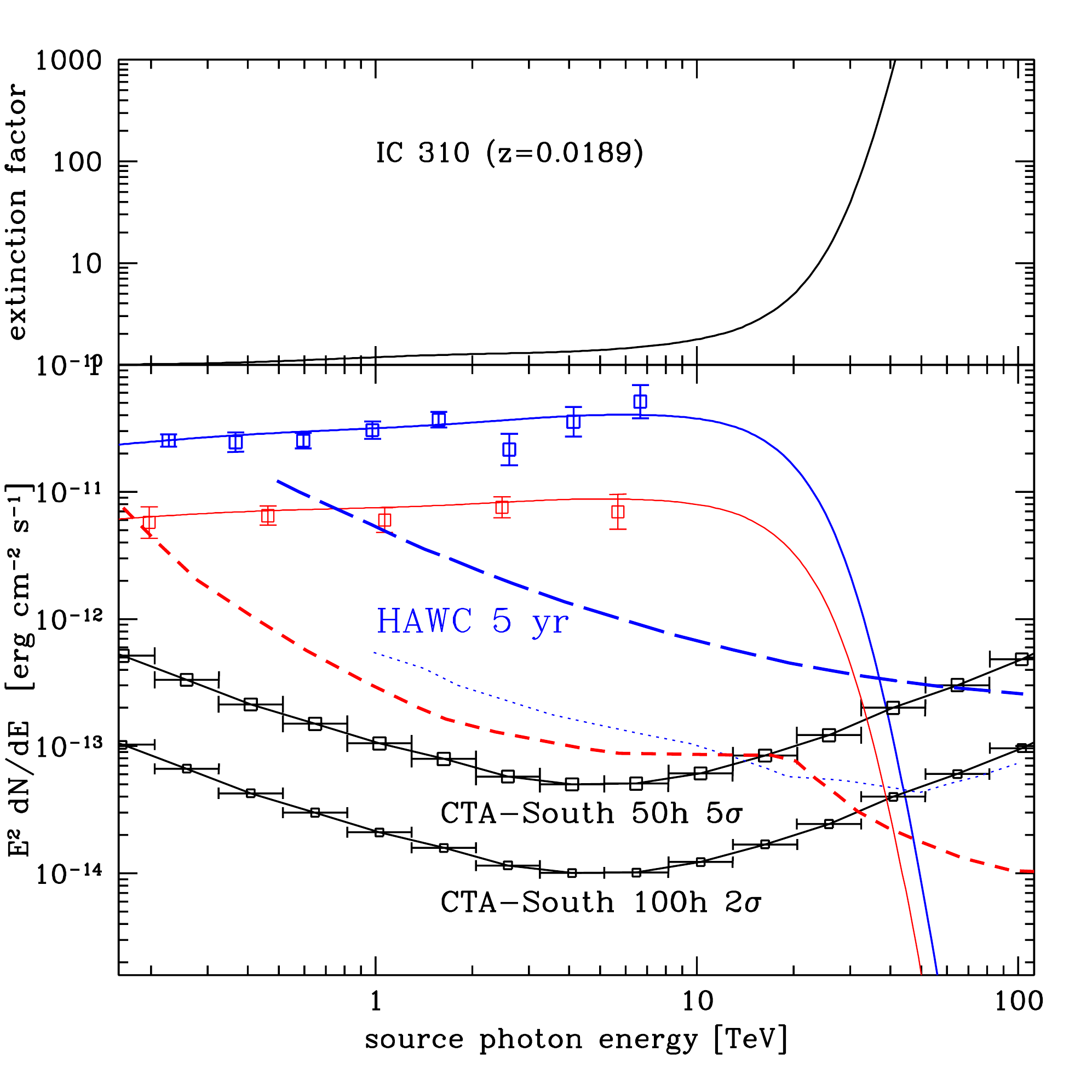}
\caption{
\textit{Top panel: } Photon--photon absorption correction ($exp[\tau_{\gamma\gamma}]$) for the source IC 310 at $z=0.0189$, based on the EBL model by FR2017.
\textit{Bottom:} The blue data-points and continuous line were taken during an outburst phase, the red data and continuous line during a prolonged high-state (see text).
The 50-hour $5\sigma$ and 100 hours $2\sigma$ sensitivity limits for CTA, and the HAWC 5 years limits are shown. The blue dotted line and the red dashed one indicate the SWGO and LHAASO 5 years $5\sigma$ limits, respectively.
}               
\label{IC310} 
   \end{figure}
%-----------------------------------------------------------------

In this section, we consider all such VHE sources in the TEVCAT catalog (\url{http://tevcat.uchicago.edu/}) closer than 100 Mpc (the distance of the two closest Markarians is about 150 Mpc). The few potentially interesting objects, for which we have simulated deep IACT observations, are discussed in the following (the coordinates refer to the J2000 calibration).

\subsection{Source IC 310}
{\it (R.A.: 03 16 43.0; Dec: +41 19 2913 25 26.4; z=0.0189).} 
This very low-redshift VHE source appears to be a transitional object between a radio galaxy and a low-luminosity HBL blazar. It shows a complex morphology, similar to a blazar but characterized by a kiloparsec-scale radio structure typical of radio galaxies. One important feature of this source is its extreme variability: during extensive campaigns in 2012 and 2013 it showed a huge flare (November 2012) and then a high state during several of the following months. Spectral data taken by MAGIC during the flare \citep{Ahnen} are reported in Fig. \ref{IC310} as blue datapoints: the spectrum is very hard (photon spectral index $\Gamma=1.9$). 
During the period 2009-10, the source displayed a high state, still very hard ($\Gamma \simeq 2$) but lasting longer and at lower brightness, as also reported by \cite{Ahnen}. 
Because of their hardness and spectral shape, we fitted the two data sets with simple power laws and corrected them for EBL absorption assuming the heuristic model by FR2017. 
These are also shown with the appropriate colors, and reported in the figure.

%-----------------------------------------------------------------
%                                                One column figure
%----------------------------------------------------------------- 
   \begin{figure}
   \centering
   \includegraphics[width=9cm]{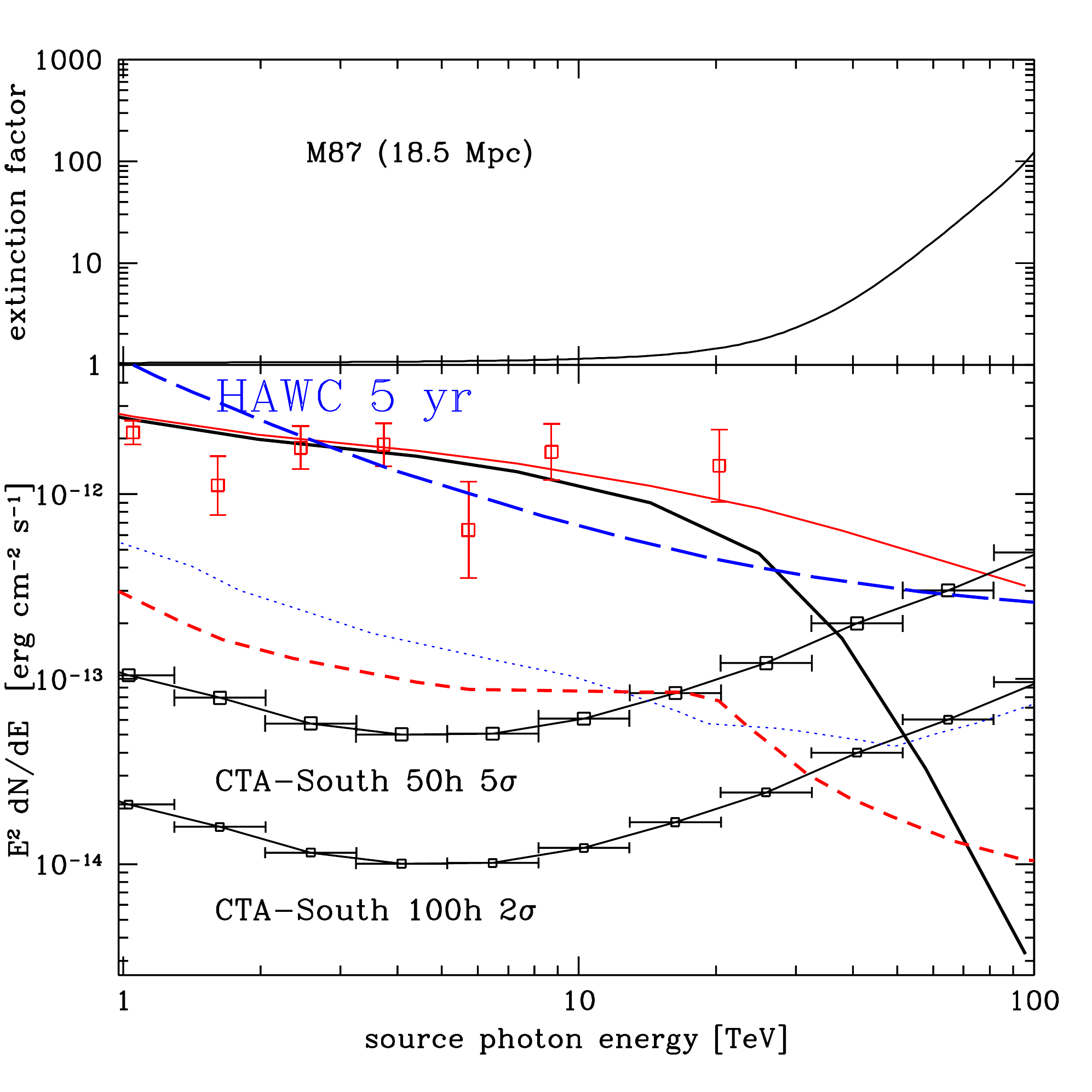}
\caption{
\textit{Top panel: }  Photon--photon absorption correction ($exp[\tau_{\gamma\gamma}]$)
for the source M 87 at 18.5 Mpc, based on the EBL model by FR2017. 
\textit{Bottom}: Observed (open red and continuous line) and absorption-corrected (black line) spectral data are reported. 
The 50-hour $5\sigma$ and 100-hour $2\sigma$ sensitivity limits for CTA, and the HAWC 5-year limit are shown. The blue dotted line and the red dashed line indicate the SWGO and LHAASO 5-year $5\sigma$ limits, respectively.
}               
\label{M87}
   \end{figure}
%-----------------------------------------------------------------

\subsection{Source  M 87}

{\it (R.A.:     12 30 47.2; Dec.:       +12 23 51; 18.5 Mpc).}  This very famous type-I Fanaroff-Riley radio-galaxy, part of the Virgo cluster, has been observed and detected by all major Cherenkov observatories (HEGRA, HESS, VERITAS, MAGIC).
The VHE source showed strong variability in intensity, but no significant spectral changes, with typically hard spectra \citep[$\Gamma=2.2$, ][]{Aharonian2006}. We report the HESS data taken during a high state in 2005 in Fig. \ref{M87}, a high state that lasted for a few months early in that year. The combination of HESS and \textit{Fermi} data reveals a clear spectral hardening above 100 GeV.

To fit and extrapolate these spectral-intensity data to the highest VHE energies, we referred to the interpretative analysis by Fraija \& Marinelli (2018). The complex broad-band spectrum is fit with a combination of a synchrotron self-Compton emission reproducing the whole spectral energy distribution from the radio to the high energies, and a hadronic component needed to fit the VHE excess. The latter would imply the acceleration of protons and production of neutral pions $\pi^0$ by interaction with ambient material or photons. 
The red line in Fig. \ref{M87} corresponds to their spectral solution slightly rescaled to best fit the HESS data during the high state.
We then corrected this rest-frame spectrum by EBL absorption, and reported it in the figure as the black continuous line.

%-----------------------------------------------------------------
%                                                One column figure
%----------------------------------------------------------------- 
   \begin{figure}
   \centering
   \includegraphics[width=9cm]{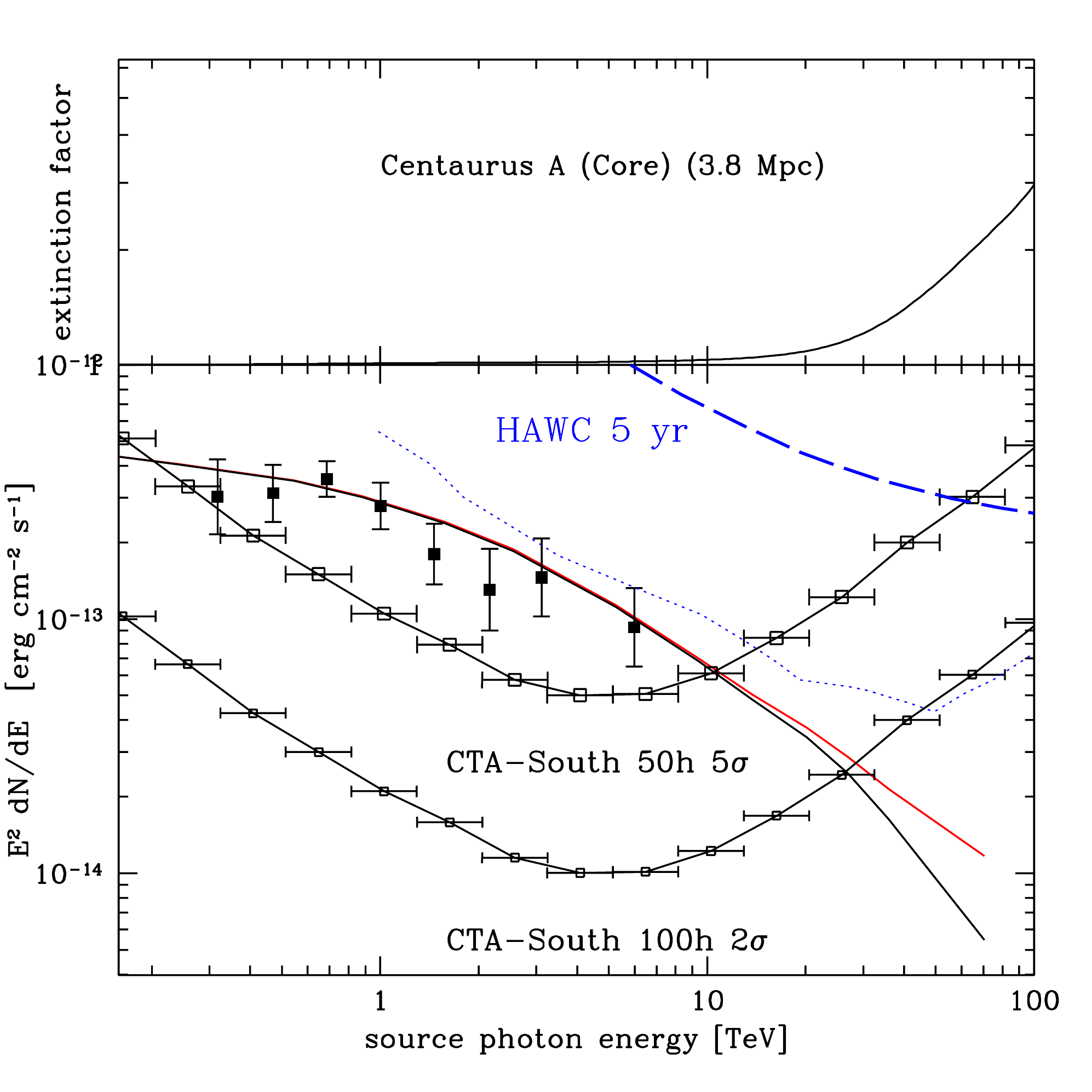}
\caption{
\textit{Top panel: } Photon--photon absorption correction ($exp[\tau_{\gamma\gamma}]$)
for the source Centaurus A at 3.8 Mpc, based on the EBL model by
FR2017. 
\textit{Bottom panel:} Observed (filled black squares) and model interpolated intrinsic spectrum (red line), and the absorption-corrected (black line) spectrum are reported. The two sensitivity limits for CTA are shown; see caption to Fig. \ref{M87}
}               
\label{CenA} 
   \end{figure}
%-----------------------------------------------------------------

\subsection{The Centaurus A core}
{\it  (R.A.:    13 25 26.4; Dec. -43 00 42; 3.8 Mpc).} This is by far the closest (3.8 Mpc) AGN, which was found to be a weak VHE emitter by the HESS observatory \citep[][among others]{Aharonian2009}. 

The source is also classified as a type-I Fanaroff-Riley radio galaxy, and includes a parsec-scale jet and counter-jet system, a kiloparsec-scale jet and inner lobes, and giant radio-emitting lobes on a scale of hundreds of kiloparsecs.
The region containing the radio core and the jets was observed with HESS from 0.3 to 10 TeV during a first long campaign from 2004 to 2008 and a second one from 2008 to 2010. 
Remarkably, over such a long sampling period of 6 years, no significant spectral or intensity variations were found, in particular none between the two cumulated datasets. The combined data based on the summed counts (for a total observing time of 213 hours) are reported as the black data points in Fig. \ref{CenA}, as published in \cite{Abdalla}.

A comparison of roughly simultaneous HESS and \textit{Fermi} data of the source revealed a similarly complex spectral energy distribution to that of M 87, with a large portion of it well reproduced by a synchrotron self-Compton model, and a similar excess above 100 GeV.
To simulate observations at the highest energies, we fitted and extrapolated the HESS spectral data with a similar model by \cite{Fraija} previously adopted for M87. In this case the VHE excess is also modeled as a hadronic process. 
The red line in Fig. \ref{CenA} is this best-fit including an extrapolation up to 100 TeV. The black line shows the EBL corrected spectrum.

\section{ Constraints on IR extragalactic background light from VHE observations of local AGNs}

Figures \ref{IC310} to \ref{CenA} report the predicted sensitivity limits for future long  observations of the three most promising local VHE-emitting AGNs. 
The three plots reveal different opportunities for constraining the IR EBL using deep observations of the three sources.

\begin{enumerate}

\item {\bf Centaurus A core}. 
Because of its proximity, this might be expected to offer, in principle, the extragalactic power source detectable up to the highest photon energies, ideally up to $\sim 100 $ TeV.  This would therefore allow us to constrain the IR EBL at the longest wavelengths of about $\lambda \sim 100\ \mu$m, approximately the expected wavelength of peak intensity for the IR EBL. This would therefore potentially be the most interesting case. 

Unfortunately, the source is very faint and has a soft spectrum at multi-TeV energies. %\LEt{Please check that I have retained your intended meaning.} With 213 hours of total exposure time, HESS detected it to 6 TeV. 
A fifty-hour observation with the CTA South array would measure the spectrum at 5$\sigma$ only up to 10 TeV, while an observation of 100 hours would detect it to about 30 TeV or so, adopting the 2$\sigma$ criterion. However, as shown in Fig. \ref{CenA}, at that photon energy the IR EBL absorption effect would be negligible, unless the real EBL intensity is much larger than that modeled by FR2017.

Apart from being in the wrong hemisphere, the HAWC five-year sensitivity is one order of magnitude shallower than the expected signal from the core of Centaurus A. A better option is offered by the  SWGO after 5 years of operation.
In any case, Centaurus A does not appear to be adequate for observing the highest VHE energies from an extragalactic source.

\item {\bf IC 310}. 
The high-declination, quickly varying source IC 310 is visible by the CTA North array, but possibly also from the south. If observed in the high state (red line spectrum), it will be detected by CTA up to 40 TeV with 50-100 hours of total integration. 
HAWC would detect it to about 30 TeV after 5 years of operation, and LHAASO again up to about 40 TeV. The source distance and consequent extremely steep spectrum at high energies make these instruments comparable to each other.

\item {\bf M 87}. 
Among all those that we have considered so far, this VHE source appears to have almost all the properties required to function as a powerful `lighthouse', illuminating the local VHE universe.  
It is a low(positive)-declination object observable from both northern and southern Cherenkov observatories, particularly the two CTA arrays, HAWC and LHAASO, and potentially SWGO.
Combined data from long observations by all of these instruments will  achieve sensitivities of a few times $10^{-14}$  erg cm$^{-2}$ $s^{-1}$. From Fig. \ref{M87} and Eq. \ref{energy}, this will allow measurements of the M 87 spectrum up to 60-70 TeV and the IR EBL intensity up to the wavelengths of $\lambda \simeq 70-90\ \mu$m, which is close to the expected IR EBL maximum.

Overall, long VHE observations of M 87 will offer the best opportunity to significantly constrain the diffuse photon density at FIR wavelengths. This analysis will require the combination of long VHE monitoring of the source with a spectral modeling of the possible photon--photon absorption by radiation fields intrinsic to the object and particle energy-loss effects.
The nucleus of M 87, because of its low bolometric luminosity, has been found to be effectively transparent for gamma rays up to energies of $\sim$10 TeV, making this source an ideal laboratory for IR EBL studies, as well as for studies of particle acceleration processes in the BH vicinity \citep{Neronov}.  
However, attempts to constrain the IR EBL intensity from M 87 VHE monitoring will go hand in hand with the necessary modeling of the nuclear intrinsic spectrum.

%-----------------------------------------------------------------
%                                                One column figure
%----------------------------------------------------------------- 
   \begin{figure}
   \centering
   \includegraphics[width=9cm]{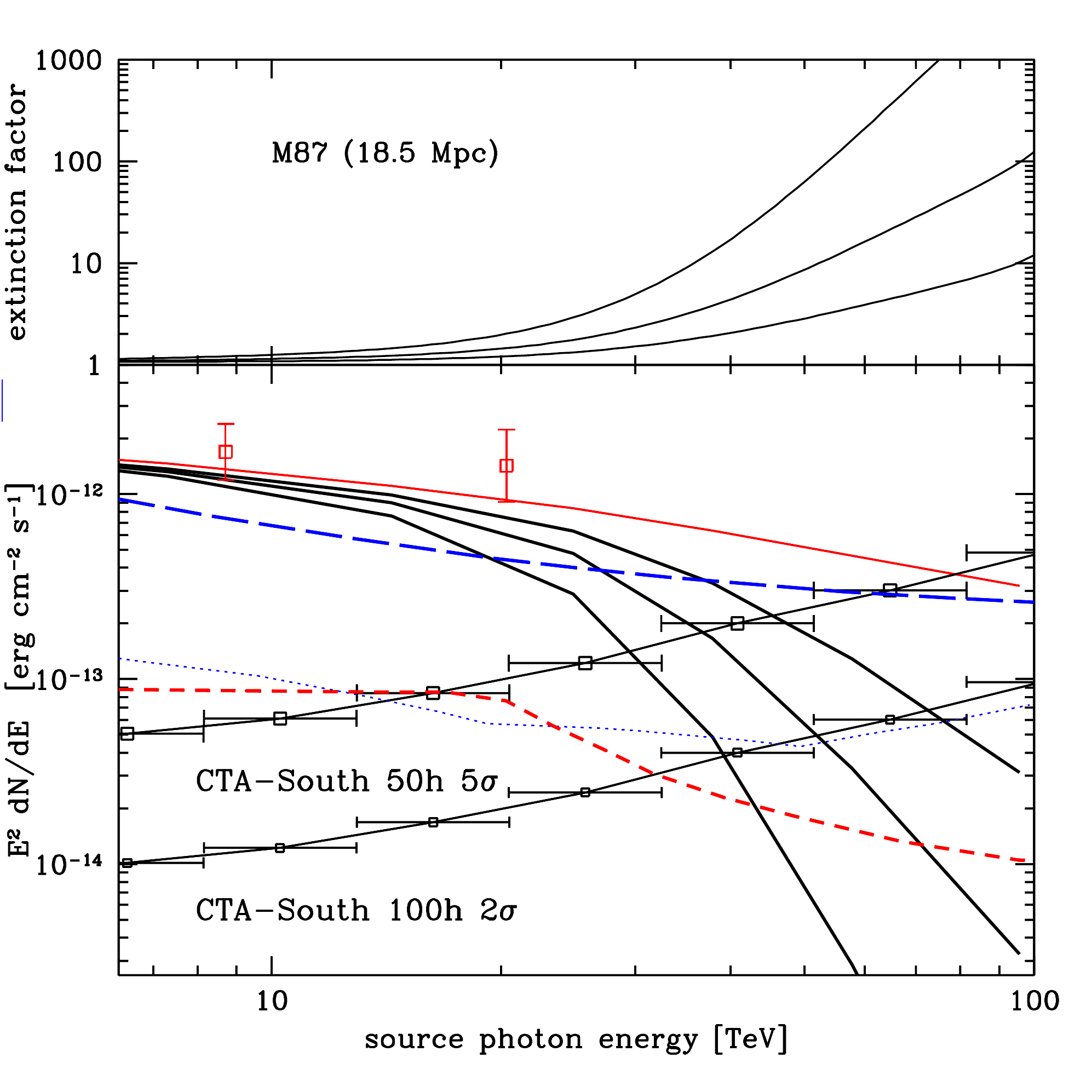}
\caption{
\textit{Top panel: } Photon--photon absorption correction to be applied to the M87 spectrum for three IR EBL adopted intensities. The intermediate curve corresponds to the FR2017, and the upper and lower curves to IR EBL intensities a factor two higher and lower between 10 and 100 $\mu$m. 
\textit{Bottom panel:} As in Fig. \ref{M87}, for the three IR EBL models.
The 50-hour $5\sigma$ and 100-hour $2\sigma$ sensitivity limits for CTA, and the HAWC and LHAASO five-year $5\sigma$ limits are also indicated.
}               
\label{M87multi}
   \end{figure}
%-----------------------------------------------------------------

An illustration of the relation between IR EBL intensity and the M87 VHE spectral convergence is given in Fig. \ref{M87multi}. Variations of a factor two in the EBL intensity with respect to the best-guess EBL model by FR2017 and source emission spectrum by \cite{Fraija} produce changes in the VHE cutoff by similar factors, which will be easily measured by CTA and perhaps even better by LHAASO.

   \end{enumerate}

\section{Conclusions}

Sources emitting VHE photons, such as AGNs and {blazars} in particular, can be used as distant lighthouses, irradiating the universe and allowing us to determine the intensity of diffuse radiation along their line-of-sight, taking advantage of the photon--photon interaction.
The effect of such interaction manifests itself in a fast exponential drop off of the source spectra at the high energies that depends on the product of the source distance and the background photon number density.

One of the least well-known diffuse background radiations is the IR EBL over the wavelength range of 10 to 100 $\mu$m, including a substantial fraction of the total light emitted by cosmic sources, particularly during ancient phases of galaxy formation. Here, direct measurements of the photon background are prevented by the huge foregrounds from IPD and Galactic dust emissions.

Constraining the IR EBL around its peak wavelength at $\sim 100\mu$m therefore requires a delicate choice between the source distance, luminosity, and intrinsic spectral shape. We investigated various kinds of low-redshift objects as prominent VHE emitters. All {blazars}, including the EHBL category, are not suited to our purpose because they are too distant. This holds true independently of the source luminosity, including extreme flaring states of nearby Markarians. 

We alternatively considered opportunities offered by more local AGNs emitting at VHE, in particular local radio-galaxies. Among the best known of these, M 87 appears to offer an ideal combination of properties, potentially allowing us to probe the IR EBL up to $\sim 100\mu$m. This will require extensive monitoring of the source with state-of-the-art observatories,  the best opportunities being offered by CTA-South, particularly during high-emission states.

\begin{acknowledgements}
      This work was supported by the University of Padova via various contracts, and by the Italian National Institute for Astrophysics (INAF). 
      The authors acknowledge contribution from the grant CTA-SKA, \textit{Probing particle acceleration and gamma-ray propagation with CTA and its precursors} (PI F. Tavecchio) of INAF.
\end{acknowledgements}

% WARNING
%-------------------------------------------------------------------
% Please note that we have included the references to the file aa.dem in
% order to compile it, but we ask you to:
%
% - use BibTeX with the regular commands:
%   \bibliographystyle{aa} % style aa.bst
%   \bibliography{Yourfile} % your references Yourfile.bib

\begin{thebibliography}{}

	 \bibitem[\protect\citeauthoryear{}{}]{}



	 \bibitem[\protect\citeauthoryear{Abdalla et al.}{2019}]{Abdalla2019}
Abdalla, H., Aharonian, F., Ait Benkhali, F., 2019, ApJ 870, 93

	 \bibitem[\protect\citeauthoryear{Abdalla et al.}{2018}]{Abdalla}
Abdalla, H., Abramowski, A., Aharonian, F., et al., 2018 A\&A...619A..71H

    \bibitem[\protect\citeauthoryear{Aharonian et al.}{1999}]{Aharonian1999}
Aharonian, F., et al., 2006, Nature, 440, 1018

 \bibitem[\protect\citeauthoryear{Aharonian et al.}{2006}]{Aharonian2006}
Aharonian, F., Akhperjanian, A. G., Bazer-Bachi, A. R., et al. 2006, Science,
314, 1424

 \bibitem[\protect\citeauthoryear{Aharonian et al.}{2007}]{Aharonian2007}
Aharonian, F., Akhperjanian, A. G., Bazer-Bachi, A. R., et al. 2006, A\&A,
475, L9

    \bibitem[\protect\citeauthoryear{Aharonian et al.}{2001}]{Aharonian2001}
Aharonian, F.A.,  Akhperjanian, A. G., Barrio, J. A., et al., 2001, A\&A 366, 62

	 \bibitem[\protect\citeauthoryear{Aharonian et al.}{2001}]{Aharonian2009}
Aharonian, F., Akhperjanian, A. G., Anton, G., et al. 2009, ApJ, 695, L40

	 \bibitem[\protect\citeauthoryear{Ahnen et al.}{2017}]{Ahnen}
Ahnen et al. 2017 2017 A\&A...603A..25A

	 \bibitem[\protect\citeauthoryear{Albert et al.}{2019}]{Albert19}
Albert, A., R. Alfaro, A., Ashkar, H., et al., 2019, arXiv:1902.08429v1

 
	 \bibitem[\protect\citeauthoryear{Antonucci \& Miller}{1985}]{Antonucci}
	Antonucci, R. R. J.; Miller, J. S., 1985, ApJ 297, 621
			
		 \bibitem[\protect\citeauthoryear{Arnouts et al.}{2005}]{Arnouts}
		Arnouts, S., Schiminovich, D., Ilbert, O., et al., 2005, ApJ 619, L43

		 \bibitem[\protect\citeauthoryear{Blain et al.}{2002}]{blain}
			Blain, A.W., Smail, I.,  Ivison, R.J.,  Kneib, J.-P., Frayer., D.T., Phys.Rept. 369 (2002) 111-176
			
	 \bibitem[\protect\citeauthoryear{Cerruti et al.}{2015}]{cerruti}
{Cerruti}, M. and {Zech}, A. and {Boisson}, C. and {Inoue}, S., 2015, MNRAS 448, 910

	 \bibitem[\protect\citeauthoryear{Costamante \& Ghisellini}{2002}]{Costamante}
Costamante L., Ghisellini G., 2002, A\&A, 384, 56

	 \bibitem[\protect\citeauthoryear{Costamante et al.}{2018}]
{Costamante18}
Costamante, L., Bonnoli, G., Tavecchio, F., Ghisellini, G., Tagliaferri, G., Khangulyan, D.,     MNRAS 477, 4257

	 \bibitem[\protect\citeauthoryear{Cucciati et al.}{2012}]{Cucciati}
	 Cucciati, O.,  Tresse, L.,  Ilbert, O., et al., 2012, A\&A 539, A31 


	 \bibitem[\protect\citeauthoryear{Daddi et al.}{2005}]{Daddi}
	Daddi, E., Renzini, A., Pirzkal, N., et al., 2005, ApJ 626, 680
	

	 \bibitem[\protect\citeauthoryear{Daddi et al.}{2007}]{Daddi2007}
	Daddi, E., Dickinson, M., Morrison, G., 	 2007, ApJ 670, 156
	
	 \bibitem[\protect\citeauthoryear{DeYoung et al.}{2012}]{DeYoung}
DeYoung, T., et al., the HAWC Collaboration,  2012, Nuclear Inst. and Methods in Physics Research, A, Volume 692, p. 72-76.


	 \bibitem[\protect\citeauthoryear{Di Sciascio et al.}{2016}]{DiSciascio}
	Di Sciascio, G., for the LHAASO Collaboration, in	Nuclear and Particle Physics Proceedings, Volume 279, p. 166-173.

			
					 \bibitem[\protect\citeauthoryear{Eales et al.}{2018}]{eales}
 Eales et al. 2018, Monthly Notices of the Royal Astronomical Society, Volume 481, Issue 1, p.1183-1194

	 \bibitem[\protect\citeauthoryear{Fermi-LAT Collaboration}{2018}]{FermiLAT}
Fermi-LAT Collaboration, Science 362, 1031–1034 (2018)

					 \bibitem[\protect\citeauthoryear{Fixten et al.}{1997}]{fixten}
		Fixen et al. (1997), ApJ 490,  2
	
	 \bibitem[\protect\citeauthoryear{Foffano et. al.}{2019}]{Foffano}
Foffano L., Prandini E., Franceschini A., Paiano S., 2019, MNRAS, 486, 1741

	 \bibitem[\protect\citeauthoryear{Fraija \& Marinelli}{2018}]{Fraija}
Fraija, N. \& Marinelli, A., 2018, preprint, ApJ in press.

	 \bibitem[\protect\citeauthoryear{Franceschini et al.}{2001}]{AF2001}
	Franceschini, A.; Aussel, H.; Cesarsky, C. J.; Elbaz, D.; Fadda, D., 2001, A\&A 378, 1
	
	 \bibitem[\protect\citeauthoryear{Franceschini et al.}{2008}]{AF2008}
	Franceschini, A., Rodighiero, G., Vaccari, M., 2008, A\&A 487, 837 (AF2008)

	 \bibitem[\protect\citeauthoryear{Franceschini \& Rodighiero}{2017}]{AF2017}
	Franceschini, A., Rodighiero, G., 2017, A\&A 603, 34  (FR2017)
	
		 \bibitem[\protect\citeauthoryear{Franceschini et al.}{2010}]{AF2010}
    Franceschini, A.; Rodighiero, G.; Vaccari, M.; Berta, S.; Marchetti, L.; Mainetti, G., Astronomy and Astrophysics, Volume 517, id.A74, 26 pp.

	 \bibitem[\protect\citeauthoryear{Fritz et al.}{2006}]{Fritz}
 Fritz, J., Franceschini, A., Hatziminaoglou, E., 2006, MNRAS366, 767	
	
	 \bibitem[\protect\citeauthoryear{Giavalisco et al.}{2004}]{Giavalisco}
	Giavalisco, M., Dickinson, M., Ferguson, H. C., et al., 	2004, ApJ 600, L103
	
					 \bibitem[\protect\citeauthoryear{Gruppioni et al.}{2013}]{Gruppioni}
Gruppioni, C., et al., Mon.Not.Roy.Astron.Soc. 432 (2013) 23

	 \bibitem[\protect\citeauthoryear{Gutierrez \& Lopez-Corredoira}{2017}]{GL}
 Gutierrez, C. M., and Lopez-Corredoira, M., ApJ 835, 111  

						 \bibitem[\protect\citeauthoryear{Hauser \& Dwek}{2001}]{hauseranddwek} 	
Hauser, M.G., Dwek, E.,	Annual Review of Astronomy and Astrophysics, Vol. 39, p. 249-307 (2001)

 \bibitem[\protect\citeauthoryear{Heitler}{1960}]{Heitler}
Heitler, W., 1960, The Quantum Theory of Radiation, Oxford Press, London

	 \bibitem[\protect\citeauthoryear{Kennicutt}{1998}]{Kennicutt}
Kennicutt, R.C., 1998, ARA\&A 36 189

			 \bibitem[\protect\citeauthoryear{Hauser et al.}{1998}]{hauser}
		Hauser et al. (1998), ApJ 508, Issue 1, 25.
	
	\bibitem[\protect\citeauthoryear{Lagache et al.}{1999}]{lagache99} 
Lagache, G., Haffner, L.M., Reynolds, R.J., Tufte,  S.L., Nov 1999, A\&A 354 (2000) 247
						
	 \bibitem[\protect\citeauthoryear{M. Longair}{2000}]{longair}
		Longair, M., High Energy Astrophysics, Cambridge University Press (2000).
	Cambridge University Press
	
 \bibitem[\protect\citeauthoryear{Madau and Dickinson}{2004}]{madau}
			Piero Madau, Mark Dickinson. Ann.Rev.Astron.Astrophys. 52 (2014) 415-486

\bibitem[\protect\citeauthoryear{Mirzoyan et al.}{2019}]{Mirzoyan}
Mirzoyan, R., Vovk, I., Peresano, M., et al., 2019, arXiv190304989M,
in Nucl. Instr. Meth. in the volume of the RICH10 conference proceedings; doi: 10.1016/j.nima.2018.11.046, 

	 \bibitem[\protect\citeauthoryear{Neronov \& Aharonian}{2007}]{Neronov}
	Neronov, A., Aharonian, Felix A., 	2007, ApJ 671, 85
	
		 \bibitem[\protect\citeauthoryear{Puget et al.}{1996}]{puget}
		Puget et al. (1996), A\&A 308, L5, 1996
	
	 \bibitem[\protect\citeauthoryear{Salpeter}{1955}]{Salpeter}
	Salpeter, E.E. 1955, ApJ  121, 161 
	
	 \bibitem[\protect\citeauthoryear{Santini et al.}{2009}]{Santini}
		Santini, P., Fontana, A., Grazian, A., et al., 2009, A\&A 504, 751
	
  \bibitem[Sarazin(1986)]{Sarazin}
	Sarazin, C., Reviews of Modern Physics, Volume 58, Issue 1, January 1986, pp.1-115
  
	 \bibitem[\protect\citeauthoryear{Shimasaku et al.}{2005}]{Shimazaku}
	Shimasaku, K., Ouchi, M., Furusawa, H., et al., 2005, Publ. Astron. Soc. Japan 57, 447
	
	 \bibitem[\protect\citeauthoryear{Schiminovich et al.}{2005}]{Schiminovich}
Schiminovich, D., Ilbert, O., Arnouts, S., et al., 2005, ApJ 619, L47

		 \bibitem[\protect\citeauthoryear{Soifer \& Neugebauer}{1991}]{Soifer}
Soifer, B.T., Neugebauer, G., 1991 AJ 101, 354S


	 \bibitem[\protect\citeauthoryear{Stanev \& Franceschini}{}]{Stanev}
Stanev, T., Franceschini, A.,  1998, ApJ 494, L 159

\bibitem[\protect\citeauthoryear{Stecker et al.}{1992}]{Stecker}
Stecker, F. W., de Jager, O. C., \& Salamon, M. H. 1992, ApJ 390, L49

	 \bibitem[\protect\citeauthoryear{Tavecchio et al.}{2009}]{Tavecchio09}
Tavecchio F., Ghisellini G., Ghirlanda G., Costamante L.,
Franceschini A., 2009, MNRAS, 399, L59

	 \bibitem[\protect\citeauthoryear{Tavecchio et al.}{2010}]{Tavecchio10}
Tavecchio F., Ghisellini G., Ghirlanda G., Foschini L., Maraschi
L., 2010, Mon. Not. Roy. Astron. Soc., 401, 1570


	 \bibitem[\protect\citeauthoryear{Thacker et al.}{2015}]{Thacker}
Thacker, C., Gong, Y., Cooray, A., De Bernardis, F., Smidt, J., and
 Mitchell-Wynne, K., ApJ 811, 125
			
	 \bibitem[\protect\citeauthoryear{Stanev and Franceschini}{1997}]{StanevandFranceschini1997}
	Todor Stanev, Alberto Franceschini Astrophys. J. 494, L159-L162 (1998)
			
	 \bibitem[\protect\citeauthoryear{Urry and Padovani}{1995}]{UrryPadovani}
Urry, M., Padovani, P.,  PASP 107, 803
			
		 \bibitem[\protect\citeauthoryear{Wyder et al.}{2005}]{Wyder}
		Wyder, T., et al.\ 2005, ApJ 619, L15
			
	 \bibitem[\protect\citeauthoryear{Zemcov et al.}{2014}]{Zemcov14}
	Zemcov, Michael; Smidt, Joseph; Arai, Toshiaki; et al., Science, Volume 346, Issue 6210, pp. 732-735 (2014)
	
	 \bibitem[\protect\citeauthoryear{Zemcov et al.}{2013}]{Zemcov13}
		Zemcov, M., et al., Astrophys. J., Suppl. Ser. 207, 31 (2013).
		
			
\end{thebibliography}
%
% - join the .bib files when you upload your source files
%-------------------------------------------------------------------

\end{document}